# Hybrid magneto-dynamical modes in a single magnetostrictive nanomagnet on a piezoelectric substrate arising from magneto-elastic modulation of precessional dynamics

Sucheta Mondal[1], Md Ahsanul Abeed[2], Koustuv Dutta[1], Anulekha De[1], Sourav Sahoo[1], Anjan Barman[1,*] and Supriyo Bandyopadhyay[2,**]

[1]Department of Condensed Matter Physics and Material Sciences, S. N. Bose National Centre for Basic Sciences, Block JD, Sector III, Salt Lake, Kolkata 700 106, India

[2]Department of Electrical and Computer Engineering, Virginia Commonwealth University, Richmond, VA 23284, USA

**Abstract**

Magneto-elastic (or "straintronic") switching has emerged as an extremely energy-efficient mechanism for switching the magnetization of magnetostrictive nanomagnets in magnetic memory, logic and non-Boolean circuits. Here, we investigate the ultrafast magneto-dynamics associated with straintronic switching in a single quasi-elliptical magnetostrictive Co nanomagnet deposited on a piezoelectric $Pb(Mg_{1/3}Nb_{2/3})O_3$-$PbTiO_3$ (PMN-PT) substrate using time-resolved magneto-optical Kerr effect (TR-MOKE) measurements. The pulsed laser pump beam in the TR-MOKE plays a dual role: it causes precession of the nanomagnet's magnetization about an applied bias magnetic field and it also generates surface acoustic waves (SAWs) in the piezoelectric substrate that produce periodic strains in the magnetostrictive nanomagnet and modulate the precessional dynamics. This modulation gives rise to intriguing hybrid magneto-dynamical modes in the nanomagnet, with rich spin wave texture. The characteristic frequencies of these modes are 5-15 GHz, indicating that strain can affect magnetization in a magnetostrictive nanomagnet in time scales much smaller than 1 ns (~100 ps). This can enable ~10 GHz-range magneto-elastic nano-oscillators that are actuated by strain instead of a spin-polarized current, as well as ultrafast magneto-electric generation of spin waves for magnonic logic circuits, holograms, etc.

## 1. INTRODUCTION

Nanomagnetic switches are potential replacements for the celebrated transistor in computing and signal processing hardware because they are energy-efficient switches that are also non-volatile. Nanomagnets exhibit interesting spin configurations such as quasi-single domain configurations, vortices, skyrmions, and magnetic monopole defects[1-3]. A wide variety of dynamics can occur in them over a broad time scale, e. g. ultrafast demagnetization, re-magnetization, precession, damping, motion, and domain wall movement[4] – all of which have their own intriguing applications[5, 6].

A nanomagnet's magnetization can be switched with an external agent in a variety of ways, among which an extremely energy-efficient approach is "straintronics"[7-9]. Here, the magnetization of a magnetostrictive nanomagnet is switched with mechanical strain (Villari effect) via elastic coupling to an underlying piezoelectric substrate that is activated by a voltage[7-11]. While this methodology has attracted attention because of its excellent energy-efficiency, to our knowledge, there has been no experimental study of the ultrafast magnetization dynamics associated with straintronic switching, even though non-straintronic magnetization reversal and dynamics of large arrays of nanomagnets[12], or even single nanomagnets, have been extensively studied[13-15]. Here, we report the study of ultrafast *strain-modulated* magnetization dynamics in a single magnetostrictive Co nanomagnet fabricated on a (001) PMN-PT piezoelectric substrate, performed using time-resolved magneto-optical Kerr effect (TR-MOKE) measurements. These studies elucidate the temporal dynamics of strain-induced magnetization rotation and reveal the time-scales associated with switching. In our study, we chose Co as the magnetostrictive nanomagnet, despite its relatively weak magnetostriction, since it is a single element and not an alloy. Alloys such as GaFe and Terfenol-D have much higher magnetostriction than Co, but they have multiple phases (not all of which produce high magnetostriction) and also pinning sites that pin the magnetization and inhibit magnetization dynamics[16]. For these reasons, Co is better suited to this study.

In our TR-MOKE setup, a femtosecond laser pump beam excites the magnetization of the Co nanomagnet to precess about an applied bias magnetic field. At the same time, the alternating electric field in that same beam also generates periodic (compressive and tensile) strain in the PMN-PT substrate from $d_{33}$ and/or $d_{31}$ coupling. This happens because the laser electric field periodically reconfigures the charge distribution on the surface of the PMN-PT substrate and that, in turn, modulates the electric field within the substrate via the Poisson equation. Since PMN-PT is piezoelectric and has been poled, the periodically modulated electric field within the substrate will produce a periodic strain due to $d_{33}$ and $d_{31}$ coupling. The strain will alternate between tensile and compressive (strain is tensile if the electric field in the substrate is in the same direction as poling and compressive if the electric field is opposite to the direction of poling). Additional periodic strain is generated in the substrate (underneath the nanomagnet) from the differential thermal expansions of the nanomagnet and the substrate due to periodic heating by the pulsed pump beam[16-18]. This thermally generated strain is however always tensile in the substrate in our experiment (its magnitude varies periodically, but the sign does not change) because the thermal coefficient of expansion of Co ($13 \times 10^{-6}$/K) is greater than that of PMN-PT ($9.5 \times 10^{-6}$/K). Note that the former mechanism requires a piezoelectric substrate while the latter does not. The latter mechanism has been studied in some earlier work[17-20], but the former has not. In our system, the former is expected to be dominant (see Supporting Information) and hence, to our knowledge, this is the first study of this effect.

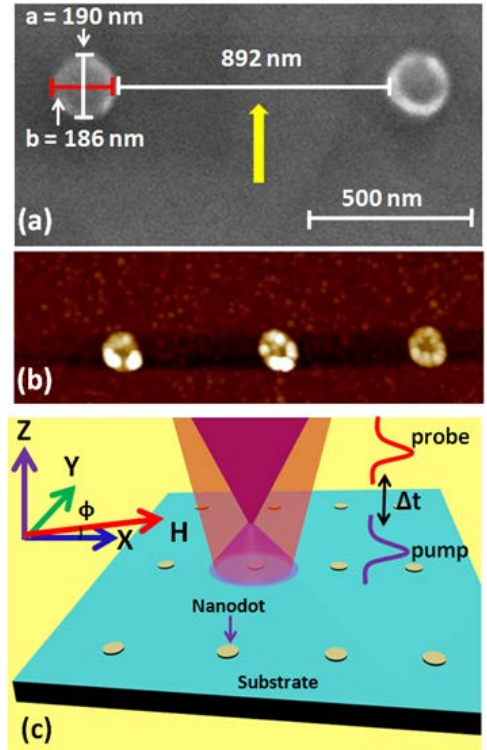

**Figure 1**: (a) Scanning electron micrograph of the Co nanomagnets deposited on a PMN-PT substrate. The edge-to-edge separation between two neighboring nanomagnets in a row [along the line collinear with their minor axis (*x*-direction)] is about 892 nm, which shows that the pitch of the array is about 1 µm. The separation between two adjacent rows is ~4 µm along the *y*-direction. Major and minor axes are denoted as *a* (≈ 190 nm) and *b* (≈ 186 nm), respectively. The yellow arrow indicates the poling direction of the substrate. (b) Magnetic force micrograph of the nanomagnets which do not show good phase contrast because of insufficient shape anisotropy. (c) The experimental geometry is shown with the bias magnetic field (*H*) applied along the array in the direction of the nanomagnets' minor axes (*x*-direction) with a slight out-of-plane tilt (ϕ) of few degrees.

The total periodic strain produced in the substrate generates surface acoustic waves (SAWs)[17-20] which periodically expand and contract the nanomagnet sitting on the substrate and change its magnetization owing to the Villari effect[21, 22]. Thus, two distinct sources induce oscillations in the out-of-plane component of the nanomagnet's magnetization: the Villari effect associated with the SAW and the Larmor precession about the bias magnetic field that can be set off by the laser-induced optical pumping. The interaction of these two oscillations gives rise to multiple hybrid oscillation modes (each with its own characteristic frequency) in the out-of-plane component of the magnetization, which then induce corresponding oscillations in the nanomagnet's reflectivity and polarization of the reflected light (Kerr signal). The periods of these oscillations are found to be 70-200 ps, which suggests that the magnetization of the nanomagnet can respond to SAW-induced strain in time scales on the order of 100 ps. This is an important insight since SAW-based magnetization switching has attracted significant interest owing to its many applications[23-38].

Our study also revealed that the spin waves which are excited in the nanomagnet in the form of hybrid magneto-dynamical modes exhibit complex power and phase profiles owing to the mixing of the precessional magneto-dynamics and SAW-induced magneto-dynamics.

To investigate the hybrid modes experimentally, we fabricated an array of slightly elliptical (eccentricity $\approx$ 1) Co nanomagnets (magnetostriction $\chi$ = 40-60 ppm) on a piezoelectric (001) PMN-PT substrate ($d_{33}$>2000 pC/N). The thickness of the nanomagnets is nominally 15 nm. Figures 1(a) and (b) show a scanning electron micrograph (SEM) of the nanomagnets and their magnetic force microscope (MFM) images. The nanomagnets are almost circular with major axis of 190 nm and minor axis of 186 nm. Because of this small aspect ratio, they do not show good phase contrast in the MFM images or single domain behavior, but because their shape anisotropy is small, the magnetic anisotropy is dominated by strain anisotropy in the presence of the SAWs. The spacing between neighboring nanomagnets (~1 µm) is large enough for the dipole interaction between them to be negligible, which means that magnetically isolated single nanomagnets are probed.

## 2. EXPERIMENTAL SECTION

To ascertain that the pump beam indeed generates SAWs in the PMN-PT substrate, we first measured the polarization and intensity of light reflected from the bare PMN-PT substrate as a function of the delay between the pump and the probe by focusing both pump and probe beams on to the bare substrate. Clear oscillations are observed in both polarization and intensity of the light reflected from the bare substrate and they can only originate from the SAWs. In Fig. 2(a), we show the intensity (reflectivity) oscillation for 15 mJ/cm$^2$ pump fluence (wavelength = 400 nm), whereas the oscillation in the polarization (Kerr signal) is shown in Fig. S1 of the Supporting Information. As expected, there are multiple oscillation modes in both reflectivity and polarization, each with a different frequency, because of the excitation of SAWs with multiple frequencies. Their frequencies at any given fluence were found to be independent of the bias magnetic field, showing that these oscillations are not of magnetic origin. They arise from the SAWs which cause periodic atomic displacements on the substrate's surface thereby changing the surface charge polarization periodically (PMN-PT is a polar material) and giving rise to polar phonons. The fast Fourier transformation (FFT) of these oscillations reveals the dominance of two peaks (2 GHz and 8 GHz) in the spectra (see Fig. 2(b)), which we conclude are the two dominant SAW frequencies excited in the PMN-PT substrate by the pulsed pump beam. There is also a small peak in the spectrum at ~16 GHz, which we ignore (it may be a second harmonic of the 8 GHz oscillation). The SAW wavelength $\lambda$ associated with the 2 GHz frequency is ascertained from the relation $v = \lambda f$ where $v$ is the phase velocity of the SAW and $f$ is the frequency. The phase velocity of SAW in a (001) PMN-PT crystal depends on the propagation direction, but it is on the order of 2000 m/sec [39]. The

2 GHz mode therefore corresponds to a wavelength of roughly 1 µm, which happens to be the pitch of the array. It therefore appears that this mode is a resonant mode determined by the geometry of the array, as in refs. [17-19]. The 8 GHz mode cannot be related to any obvious geometric feature of the nanomagnet array and may be intrinsic to the substrate.

We next probe the out-of-plane magnetization dynamics (temporal variation in the spatially averaged out-of-plane component of the magnetization) of a single Co nanomagnet on the PMN-PT substrate in the presence of both a bias magnetic field and SAWs by focusing the pump and probe beams (probe beam wavelength = 800 nm and fluence = 2 mJ/cm$^2$) on a *single* nanomagnet and measuring the Kerr oscillations as well as oscillations in the reflectivity. We were able to obtain signals from a single nanomagnet without cavity enhancement. In Figs. 3 (a) and (b), we show the background-subtracted time-resolved Kerr oscillations in time as a function of bias magnetic field along with their Fourier transforms. The pump fluence was 15 mJ/cm$^2$. The reflectivity oscillations and their Fourier transforms are shown in Fig. S2 of the Supporting Information.

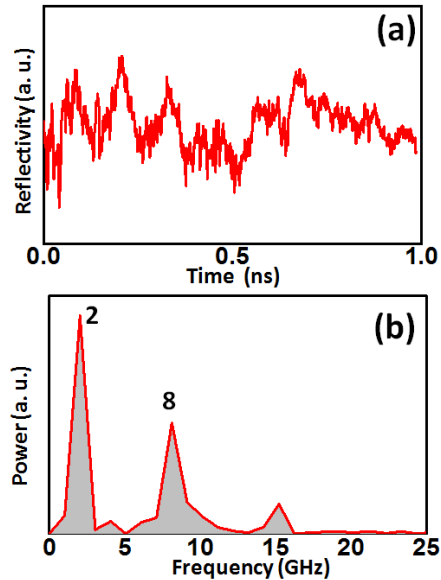

**Figure 2**: (a) Background subtracted time-resolved data for reflectivity of the bare PMN-PT substrate as a function of the delay between the pump and the probe, obtained at 15 mJ/cm$^2$ pump fluence. (b) Also shown are the fast Fourier transforms of the oscillations. Frequencies of the two most intense peaks are indicated in GHz.

Multiple modes (corresponding to peaks in the Fourier spectra) are observed in both Kerr and reflectivity oscillations at all bias magnetic fields used, with all peaks shifting to lower frequencies with decreasing bias field. This magnetic field dependence shows that these modes have a magnetic component mixed in. The most intense peak (at all bias fields except 700 Oe) is denoted as 'F' in Fig. 3(b). Two other prominent peaks that appear at higher and lower frequencies with respect to F are indicated as '$F_H$' and '$F_L$', respectively. All modes associated with these three peaks in the Fourier spectra are hybrid magneto-dynamical modes arising from the interaction between the magnetic (precession of the magnetization about the bias magnetic field) and the non-magnetic (periodic change in the magnetization due to the periodic strain generated by the SAWs) dynamics. The former dynamics depends on the bias magnetic field while the latter does not. We observed that in all cases, the amplitudes of the reflectivity oscillations are much smaller than those of the Kerr oscillations as shown in Fig. S2 of the Supporting Information.

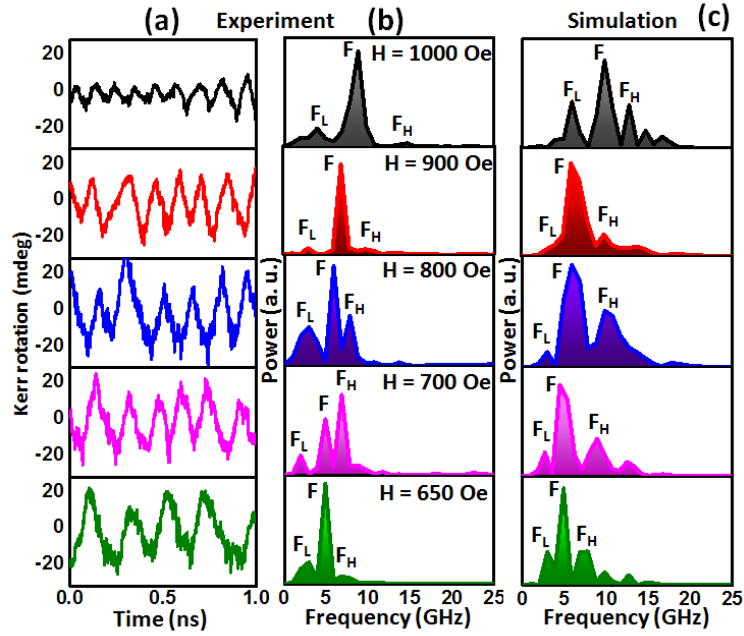

**Figure 3**: Bias magnetic field dependence of the (background-subtracted) time-resolved Kerr oscillations from a single Co nanomagnet on a PMN-PT substrate. The pump fluence is 15 mJ/cm$^2$. (a) The measured Kerr oscillations in time and (b) the fast Fourier transforms of the oscillations. The Fourier transform peaks shift to lower frequencies with decreasing bias magnetic field strength. There are multiple oscillation modes of various Fourier amplitudes. Out of those, the dominant mode (at all bias fields except 700 Oe) is denoted by F and its nearest modes $F_H$ and $F_L$. (c) Fourier transforms of the temporal evolution of the out-of-plane magnetization component at various bias magnetic fields simulated with MuMax3 where the amplitude of the periodically varying strain anisotropy energy density $K_0$ is assumed to be 22,500 J/m$^3$. The simulation has additional (weak) higher frequency peaks not observed in the experiment. The spectra in the two right panels are used to compare simulation with experiment.

## 3. RESULTS AND DISCUSSION

The Kerr and reflectivity oscillations are due to the temporal variation of the spatially-averaged out-of-plane component of the magnetization in the nanomagnet. Therefore, in order to understand the origin of the modes observed in the Kerr and reflectivity oscillations, we have modeled the magnetization dynamics in the Co nanomagnet in the presence of the bias magnetic field and the periodic strain due to the SAWs using the MuMax3 package[40]. The SAW-induced periodic strain anisotropy in the nanomagnet is included in the simulation by making the oscillating strain anisotropy energy density $K_1(t) = K_0 \left[ \sin(2\pi f_1 t) + \sin(2\pi f_2 t) \right]$, where $K_0 = (3/2)\chi\sigma$ and $f_1$, $f_2$ are the frequencies of the two dominant SAW modes (2 GHz and 8 GHz) observed for the pump fluence of 15 mJ/cm². The variation of $K_1(t)$ with time $t$ is shown in Fig. S3 of the Supporting Information. Note that the Kerr and reflectivity signals were obtained at the same fluence, which is why we consider the dominant SAW modes at that fluence. We include only two frequency components in the periodic strain anisotropy ($f_1$ = 2 GHz and $f_2$ = 8 GHz) since these two frequency components are dominant and have much higher amplitude than other components in the SAW signal (as shown in Fig. 2). Here $\chi$ is the magnetostriction coefficient of the nanomagnet (40 ppm) and $\sigma$ is the maximum stress generated in it by the SAW (stress amplitude). There are also some additional contributions to the periodic stress because of the differential thermal expansions of Co and PMN-PT due to periodic heating by the pump beam, which we neglect in this first-order analysis.

We treat $K_0$ as a fitting parameter. In the MuMax3 modeling, we generate the time-varying components of the nanomagnet's magnetization $\left[ M_x(x,y,z,t), M_y(x,y,z,t), M_z(x,y,z,t) \right]$ for different values of $K_0$. The minor axis of the nanomagnet is in the $x$-direction, the major axis is along the $y$-direction and the $z$-direction is the out-of-plane direction. We first generate the micromagnetic distribution within the nanomagnet in the presence of the bias magnetic field applied along the $x$-direction (but with no strain present) from MuMax3. Then, with this distribution as the initial state, we turn on the two-frequency oscillating strain and a small out-of-plane (square-wave) tickle pulse of amplitude 30 Oe, rise time 10 ps, and duration 100 ps to set the magneto-dynamics in motion. We find $M_x(x,y,z,t), M_y(x,y,z,t), M_z(x,y,z,t)$ for different bias magnetic fields and different values of $K_0$. We then perform fast Fourier transform of the spatially averaged value of $M_z(x,y,z,t)$, which we call $\overline{M}_z(t)$, and the spectra obtained from this transform (for different bias magnetic fields) are shown in Fig. 3 (c). They are compared with the spectra of the Kerr oscillations found experimentally and shown in Fig. 3 (b). This comparison leads us to the best fit value of $K_0$. We find reasonably good agreement between the three dominant peaks $F_L$, F and $F_H$ of the simulated Kerr spectra and the experimentally measured Kerr spectra for a value $K_0$ = 22,500 J/m³ at low bias fields, as shown in Fig. 4(a). The agreement deteriorates slightly at high bias fields, showing that perhaps a single fitting parameter $K_0$ is not adequate to model the entire range of bias fields used in the experiments.

From the value $K_0$ = 22,500 J/m³, we can also find the effective stress $\sigma$ generated in the nanomagnet. Since $K_0 = (3/2)\chi\sigma$, we find $\sigma$ = 375 MPa. We can then find the strain amplitude generated in the nanomagnet from Hooke's law: $\sigma = Y\varepsilon$, where $Y$ is the Young's modulus of Co (209 GPa)[41] and $\varepsilon$ is the strain amplitude generated by the (polychromatic) SAW. This yields $\varepsilon$ = 0.18%. Here, we treat the stress and strain as scalars since we are interested not so much in the accurate stress distribution, but rather the maximum uniaxial strain that is generated in any direction. There are reports of strain > 0.6% generated in PMN-PT[42], so, this large value of strain is not unusual. We can then calculate the effective electric field that would be required to produce the strain of 0.18% in PMN-PT. The reported $d_{33}$ values in PMN-PT are on the order of 2800 pC/N [43]. Hence the effective electric field $E$ in the PMN-PT substrate (calculated from $\varepsilon = d_{33}E$) is ~6.4 × $10^5$ V/m.

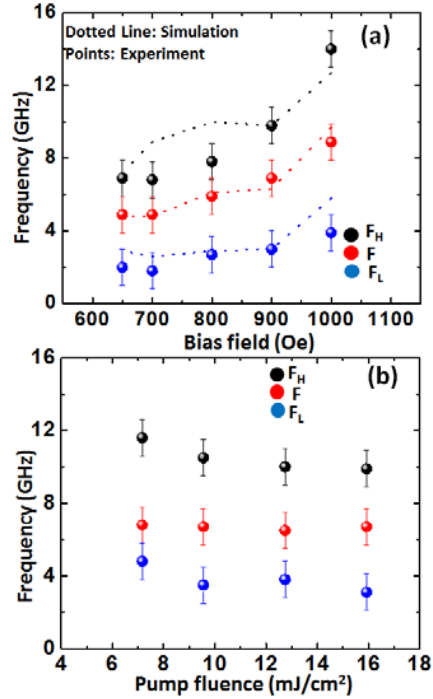

**Figure 4**: (a) Bias field ($H$) dependence of the observed three dominant frequencies in the Kerr oscillations $F_L$, F and $F_H$ (frequencies of hybrid magneto-dynamical modes) at 15 mJ/cm$^2$ pump fluence. We also show alongside (with a dotted line) the Kerr oscillation frequencies obtained from the MuMax3 modeling based on the assumption $K_0 = 22{,}500$ J/m$^3$. The intermediate points in the dotted line are a guide to the eye and are extrapolated. The simulated data points are, for the most part, within the error bars of the experimentally measured data points. (b) Fluence dependence of the frequencies at $H = 900$ Oe.

In Fig. 4(b), we show the pump fluence dependences of the three peak frequencies $F_L$, F and $F_H$ in the Kerr oscillation spectra (frequencies of the hybrid magneto-dynamical modes) for a bias magnetic field strength of 900 Oe. We expect that the frequency of the magnetic (precessional) component of the mode should depend primarily on the bias magnetic field and hence should be independent of fluence. Since the frequency of the dominant hybrid mode, F (the one with the largest Fourier component at most bias fields) is almost fluence-independent, and the frequencies of the other two modes are weakly fluence-dependent, it appears that the magnetic component of the hybrid mode (the one associated with precessional dynamics) dominates over the non-magnetic one (the one associated with SAWs) in the hybridization process. This is further supported by the observation that the mode frequencies have a pronounced bias magnetic field dependence, which would not have happened if the non-magnetic component was dominant. Note, however, that the magnetic field dependence of none of the three dominant mode frequencies ($F_L$, F or $F_H$) could be fitted with the Kittel formula[44] showing that none is a pure Kittel mode because of the hybridization.

The temporal evolutions of the micromagnetic distributions in the nanomagnet obtained from MuMax3 simulations (see Fig. S4 in the Supporting Information) reveal that in the absence of the SAWs, the bias magnetic field will tend to align the magnetization everywhere within the nanomagnet along its own direction, except perhaps at the edges. In the presence of the SAWs, the *total* field (resultant of bias magnetic field and the periodic effective magnetic field due to strain) periodically changes the orientation of the magnetization, giving rise to spin waves. This will happen even if we ignore the precession around the bias field caused by the femtosecond pump beam excitation. The time-lapsed images in Fig. S4 in the Supporting Information correspond to a number of spin wave modes superposed with appropriate powers and phases.

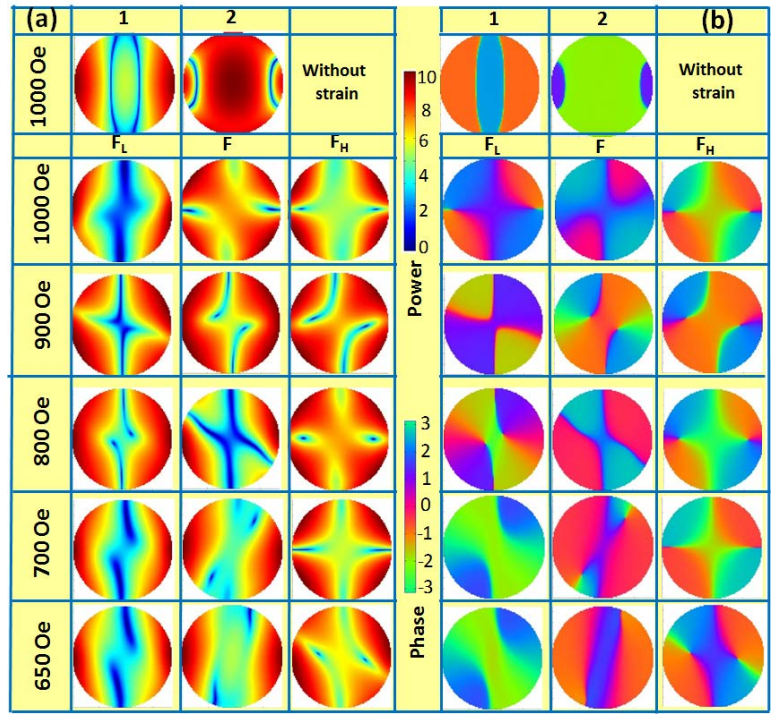

**Figure 5**: (a) Simulated power and (b) phase profiles of the spin waves associated with the three dominant frequencies $F_L$, F and $F_H$ in the Kerr oscillations at any given bias field. The top most row shows edge and center modes at the two dominant frequencies in the Kerr oscillations in the absence of strain at 1000 Oe bias field. The units of power and phase are dB and radians, respectively.

To gain more insight into the nature of the hybrid magneto-dynamical modes associated with the three dominant frequencies in the Kerr oscillations, $F_L$, $F$ and $F_H$, we have calculated the power and phase distributions of the spin waves associated with these modes, using a home-built code *Dotmag*[45]. A description of how Dotmag calculates the power and phase distributions of individual spin wave modes associated with a given mode frequency is given in the Supporting Information.

We first theoretically examine the *unstrained* nanomagnet that is not subjected to SAWs and hence experiences pure precessional dynamics because of the bias magnetic field. In this case, the fast Fourier transform of $\bar{M}_z(t)$ reveals two (bias-dependent) dominant frequencies at any given bias magnetic field (see Fig. S5(a) in Supporting Information). The power and phase distributions of the spin wave modes associated with these two frequencies are shown in the first row of Fig. 5 for a bias magnetic field of 1000 Oe applied along the ellipse's minor axis. These distributions reveal that one of the spin wave modes in the unstrained nanomagnet is a center mode (mode 2), with power concentrated in the center, and the other is an edge mode (mode 1) with power concentrated at two vertical edges which are perpendicular to the bias field[46]. The bias field dependences of the frequencies of these two modes can be found in Fig. S5(b) of the Supporting Information and can be fitted with the Kittel formula[41]. The two modes show excellent stability with bias field amplitude.

When the periodic strain anisotropy due to the SAWs (with frequencies of 2 and 8 GHz and energy density amplitude $K_0 = 22,500$ J/m$^3$) is introduced, the nature of the spin waves associated with the dominant frequencies $F_L$, $F$ and $F_H$ in the Kerr oscillations becomes very different. Instead of center and edge mode behavior, the hybrid magneto-dynamical modes of frequencies $F_L$, $F$ and $F_H$ display complex spin wave profiles with their unique characteristics. The most intense mode $F$ at 1000 Oe field contains power throughout the nanomagnet, while the phase profile suggests that spins precess alternatively in opposite phases giving rise to the azimuthal contrast with azimuthal mode quantization number $n = 5$. On the other hand, the spin waves corresponding to the two modes at frequencies $F_L$ and $F_H$ appear as the modified edge mode and quantized mode of the nanomagnet. As the bias field is reduced, the mode with frequency $F$ starts showing some quantization at the intermediate fields and finally transitions to edge mode like behavior in the low field regime. At an intermediate bias field of $H = 800$ Oe, the higher frequency mode ($F_H$) shows azimuthal character with $n = 5$. An important feature is that spin wave power is no longer concentrated along the vertical edges of the nanomagnet like in the conventional edge mode (perpendicular to the applied field) but gets slightly rotated. This is a consequence of having the SAW induced time-varying strain field present, along with the constant bias field in the system, and their competition. The component of the effective magnetic field due to strain in the *i*-th direction ($i = x, y, z$) is approximately
$H^i_{strain}(t) = 3\chi\sigma(t)m_i(t)/(\mu_0 M_s) = 2K_1(t)m_i(t)/(\mu_0 M_s)$, where $m_i(t)$ is the *i*-th component of the (spatially averaged) normalized magnetization of the nanomagnet at time $t$ [47]. The amplitude of this field is 409 Oe when $K_0 = 22,500$ J/m$^3$. Hence, it is of the same order as the bias field when it is at its maximum. The spin wave profiles in Fig. 5 reveal the rich physics of the interaction between the bias field and SAW that produces the hybrid magneto-dynamical modes.

## 4. CONCLUSION

In conclusion, we have observed hybrid magneto-dynamical modes in a single quasi-elliptical nanomagnet of Co fabricated on top of a poled (001) PMN-PT substrate due to magneto-elastic modulation of the precessional dynamics. The frequencies of these modes are in the neighborhood of 10 GHz, corresponding to ~100 ps time scale dynamics. This shows that strain can affect the magnetization state of even a weakly magnetostrictive nanomagnet in time scales far shorter than 1 ns. The spin wave textures of these modes display rich variation with the applied bias field.

Magneto-electrically generated spin waves have important applications, such as in magneto-electric spin wave amplifier for logic[48, 49]. This work sheds light on the practical operating speeds of such devices. This work also has implications for magnetic nano-oscillators which are used as nanoscale microwave generators[50], spin wave emitters, and in artificial neural networks[51]. They are usually actuated by a spin-polarized current delivering a spin-transfer-torque. One can actuate them with strain, instead of a spin -polarized current, if they are fashioned out of a magnetostrictive nanomagnet elastically coupled to an underlying piezoelectric substrate[52]. Since strain-induced switching is far more energy-efficient than spin-transfer-torque (STT) switching, such nano-oscillators will dissipate far less energy than their STT-counterparts. This work shows that the frequency range of operation of straintronic nano-oscillators can be ~10 GHz, which makes them an attractive alternative for the more common spin-torque based nano-oscillators.

Finally, there are other mechanisms at the interface of a ferromagnet and a ferroelectric that can affect the magnetic moments in the ferromagnet. One such mechanism is associated with charge screening at the interface that can affect the magnetic moments or magnetic anisotropy[53]. These mechanisms usually require an epitaxially grown thin ferromagnetic film that has strong bonding at the interface with the ferroelectric substrate. Our structures are not thin films and not epitaxially grown on the PMN-PT substrate. They are amorphous or polycrystalline Co nanostructure deposited (not epitaxially grown) on a ferroelectric substrate by electron-beam evaporation of Co through lithographically delineated windows in a resist. It is therefore very unlikely that these other mechanisms can play any significant role.

**Methods**

**Sample preparation**: The nanomagnets were fabricated with electron-beam lithography and electron beam evaporation of Co, followed by lift-off. Two different electron-beam resists of different molecular weights were used to facilitate lift off. Prior to fabrication, the PMN-PT substrate was poled with a high electric field in the direction of the major axes of the elliptical nanomagnets.

**Time-resolved magneto-optical measurements**: A two-color pump-probe technique (pump wavelength = 400 nm, pulse width = 100 fs, repetition rate = 80 MHz; probe wavelength = 800 nm, pulse width = 80 fs, repetition rate = 80 MHz) was used to measure the Kerr rotation and reflectivity signals as a function of time. The spot size of the probe beam was about 800 nm, while the pump beam was slightly defocused at the focal plane of the probe beam to obtain a spot size of about 1µm. It is very important to make the center of the pump beam coincide with that of the probe beam in order to collect the signal corresponding to uniform excitation. Since the diameter of a nanomagnet is about 190 nm and the pitch of the array is ~1 µm in a row, the probe beam can be carefully placed and maintained on top of a *single* nanomagnet using a high-precision *x-y-z* piezoelectric scanning stage with feedback loop. We were able to obtain signal from a single nanomagnet without cavity enhancement. Further measurement details can be found in the Supporting Information.

**Micromagnetic simulations**: Micromagnetic simulations are carried out with the MuMax3 and Dotmag software. We first run the MuMax3 simulation with the bias magnetic field present, but strain absent, for 1 ns to allow the micromagnetic distribution to settle to steady state. It is verified that running the simulation for a longer time does not change the distribution perceptibly. This distribution is then used as the initial state of the nanomagnet when the oscillating strain and out-of-plane tickle field are turned on in the simulation. The simulation provides the time variation of the (spatially averaged) out-of-plane component of the magnetization $\bar{M}_z(t)$ for any given bias field and oscillating strain anisotropy energy density amplitude $K_0$. Its Fourier transform spectrum corresponds to that of the Kerr oscillations at that bias field and strain amplitude. The peaks in this spectrum correspond to the (theoretically calculated) frequencies of the hybrid magneto-dynamical modes at that bias field and strain amplitude. The points in the dotted lines of Fig. 4 are found by this procedure. Note that only the points in the dotted lines at bias fields of 650, 700, 800, 900 and 1000 Oe have been computed and are meaningful. The rest of the points are extrapolations and are a guide to the eye. Further details of the simulators and simulation strategy can be found in the Supporting Information.

## ASSOCIATED CONTENT

**Supporting Information**

The Supporting Information is available free of charge at the ACS Publications website at …

Additional information on time-resolved magneto-optical measurements, micromagnetic simulations, investigation of SAW generated in the PMN-PT substrate, comparison between the amplitudes of the Kerr signal and reflectivity signals, time-varying strain used in the MuMax3 simulation, time lapsed micromagnetic distributions obtained from MuMax3 simulations, and simulated spin wave spectra and mode profiles in an unstrained nanomagnet at various bias magnetic fields.

## AUTHOR INFORMATION

Corresponding authors

[*]Email: abarman@bose.res.in

[**]Email: sbandy@vcu.edu

**Author contributions**: S. M., K. D. and A. D. made the time-resolved magneto-optical measurements. M. A. A. fabricated all the samples and characterized with magnetic force microscopy. S. M, M. A. A. and S. S. carried out the simulations. A. B. and S. B. supervised the project and formulated the problem. S. M., A. B. and S. B performed the data interpretations. All authors contributed to writing the paper.

ACKNOWLEDGEMENTS

M. A. A. and S. B. are supported by the US National Science Foundation under grant ECCS-1609303. S. B. also acknowledges the Department of Science and Technology (DST), Government of India, for the VAJRA Faculty Fellowship that enabled his summer visit to the S. N. Bose National Center for Basic Sciences in the summer of 2018. SM, AD and KD acknowledge DST, Govt. of India for INSPIRE Fellowship. S. S acknowledges the S. N. Bose National Center for Basic Sciences for financial support.

**The authors declare no competing financial interest.**

# Supporting Information

## Hybrid magneto-dynamical modes in a single magnetostrictive nanomagnet on a piezoelectric substrate arising from magneto-elastic modulation of precessional dynamics

Sucheta Mondal[1], Md Ahsanul Abeed[2], Koustuv Dutta[1], Anulekha De[1], Sourav Sahoo[1], Anjan Barman[1, *] and Supriyo Bandyopadhyay[2, *]

[1]Department of Condensed Matter Physics and Material Sciences, S. N. Bose National Centre for Basic Sciences, Block JD, Sector III, Salt Lake, Kolkata 700 106, India

[2]Department of Electrical and Computer Engineering, Virginia Commonwealth University, Richmond, VA 23284, USA

Corresponding author emails: Anjan Barman (abarman@bose.res.in); Supriyo Bandyopadhyay (sbandy@vcu.edu)

**Experimental details**

**Time-resolved magneto-optical Kerr effect measurements:** We have used a two-color pump-probe technique, where the fundamental laser beam generated from a Tsunami (Spectra Physics, λ = 800 nm, pulse width ≈ 80 fs) femtosecond laser is exploited to probe the magnetization dynamics of the sample. In all experiments, the probe fluence is kept fixed at 2 mJ/cm$^2$ and the pump fluence is varied between 7 and 15 mJ/cm$^2$. A bias magnetic field is applied at a small angle (ϕ) of about 10°-15° with respect to the sample plane (as shown in Fig. 1(c)) in the direction of the ellipses' minor axes to induce precession of the nanomagnet's magnetization around this field upon excitation by the pump beam.

To measure the time-resolved oscillations in the intensity and polarization of the light reflected from the nanomagnet, we use a 1 ns time window with 1 ps temporal resolution, which captures the full time-domain spin dynamics carrying several spin-wave frequencies. The probe beam collects information about dynamic Kerr signal and reflectivity from the sample and sends it to an optical bridge detector (OBD). A polarized beam splitter (PBS) at the entrance of the OBD splits the beam into two orthogonal polarization components. These two parts, with two different intensities (IA and IB)**,** are then detected by two photodiodes. The outputs of these two photodiodes are subsequently pre-amplified and used as inputs for two operational amplifiers to measure the total signal A+B (i.e., IA+ IB) and the difference signal A-B (i.e. IA- IB). The outputs of these two Op-Amps are measured by two lock-in-amplifiers in a phase sensitive manner with the chopper frequency as the reference. Initially, in the absence of the pump beam, the optical axis of this PBS is set at 45° to the plane of polarization of the probe beam making IA = IB, i.e., A-B = 0 and the detector is said to be in a "balanced" condition. In the next step, when the pump beam excites the nanomagnet, the plane of polarization of the probe beam is rotated due to the magneto-optical Kerr effect. Consequently, the optical axis of the PBS is no longer at 45° to the plane of polarization of the probe beam. As a result, IA ≠ IB and A-B ≠ 0. The linear magneto-optical Kerr rotation is proportional to the sample magnetization, hence to A-B. This can also provide information about the time varying polarization of the substrate. Thus, by measuring A-B as a function of time, magnetization dynamics over different time-scales are observed. The time-varying reflectivity (A+B) can provide information about charge and phonon dynamics.

**Micromagnetic Simulation:** We have performed micromagnetic simulations using the MuMax3 software. For visualization of the simulated results, we have used MuView software. In the simulation, we have considered an elliptical single nanomagnet with major and minor axes of dimensions 190 nm and 186 nm, and thickness 16 nm. The sample is discretized into cells of dimension 2 × 2 × 16 nm$^3$. The cell size in the lateral plane is kept below the exchange length of cobalt to reproduce the observed magnetization dynamics. The magnetic parameters used for the simulation are: saturation magnetization $M_s$= 1100 emu/cm$^3$, gyromagnetic ratio γ = 17.6 MHz/Oe and exchange stiffness constant $A_{ex}$= 3.0 × 10$^{-5}$ erg/cm.

In the simulation, the external bias field $H$ is first applied along the minor axis of the nanomagnet to prepare the static micromagnetic distribution by letting the simulation run for 1 ns (enough to obtain steady state). After 1 ns, the magnetization aligns along $H$ almost everywhere within the nanomagnet. Then the SAW-induced periodic strain anisotropy is introduced by making the strain anisotropy energy density $K_1(t) = K_0 \left[ \sin(2\pi f_1 t) + \sin(2\pi f_2 t) \right]$ where $K_0$ = 22500 J/m$^3$ and $f_1$, $f_2$ are the frequencies of the two dominant SAW modes (2 GHz and 8 GHz) as shown in Fig. 2 of the text. The parameter $K_0$ is used as a fitting parameter to obtain the best possible match between simulation and experiment, and the value of $K_0$ = 22500 J/m$^3$ provided the best match. In the presence of the applied bias field and the strain-generated effective field, a square pulsed magnetic field of 10 ps rise time, 100 ps width, and peak amplitude of 30

Oe is applied perpendicular to the sample plane to initiate the precession about the bias magnetic field. The resulting simulated magnetization dynamics data are acquired for 1 ns time window.

The MuMax3 simulations are run with a time step of 1 ps and hence provide information about the magnetization of the sample as a function of space (*x, y, z*) every ps. The micromagnetic distributions within the sample at different instants of time (Fig. S4) correspond to the superposition of a number of spin wave modes with varying powers and phases. The profiles in Fig. S4 do not immediately provide information about the individual resonant modes since the MuMax3 simulations yield only the spatial distribution of magnetization as a function of time: *M (t, x, y, z)*. After fixing the *z*-coordinate at a particular value $(z = z_m)$, a discrete Fourier transform (FFT) of $M(x, y, z, t)|_{z=z_m}$ with respect to time is performed to yield $\tilde{M}^{z_m}(f, x, y) = FFT\left[M(x, y, z, t)|_{z=z_m}\right]$. The frequency resolution in the FFT depends upon the total simulation time. Spatial resolution depends upon the cell size of the system adopted during the MuMax3 simulations. The fixed value of the z-coordinate $(z_m)$ is chosen to be at the surface of the nanomagnet. Our in-house software *Dotmag* then plots the $(x, y)$ spatial distribution of the power and phase of the spin waves at chosen frequencies on the surface of the nanomagnet $(z = z_m)$ according to the following relations:

$$P^{z_m, f_n}(x, y) = 20\log_{10}\tilde{M}^{z_m}(f_n, x, y);$$

$$\phi^{z_m, f_n}(x, y) = \tan^{-1}\left(\frac{\text{Im}\left(\tilde{M}^{z_m}(f_n, x, y)\right)}{\text{Re}\left(\tilde{M}^{z_m}(f_n, x, y)\right)}\right),$$

where $f_n$ is the frequency of a resonant mode.

**Investigation of surface acoustic wave (SAW) generated in the PMN-PT substrate**

The pump beam generates a SAW in the PMN-PT substrate, which gives rise to the variation in time of the polarization of light reflected from the bare PMN-PT substrate. The outputs of two photodiodes (A and B) inside an optical bridge detector provide the reflectivity signal (A+B) and the Kerr oscillation signal (A-B). The reflectivity variation in time (as a function of the delay between the pump and the probe) was shown in the main text. The time variation of the Kerr signal (polarization of reflected light) is shown in Fig. S1(a). After performing FFT of the background-subtracted time-resolved Kerr signal from the substrate, we can obtain the SAW frequencies as shown in Fig. S1(b). The two main frequencies are again 2 and 8 GHz, as in Fig. 2. Thus, the SAW frequencies obtained from the reflectivity signal and Kerr signal agree.

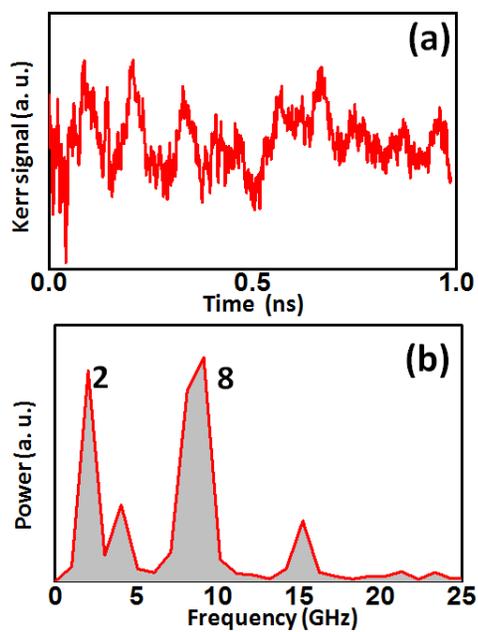

**Figure S1**: (a) Back ground subtracted time resolved data and (b) Fourier transform (or frequency spectra) of the oscillations in the polarization of light reflected from the bare PMN-PT substrate. The frequencies of the two most intense peaks are indicated in GHz and are the same as in Fig. 2.

# Comparison between the amplitudes of the Kerr signal and the reflectivity signal obtained from the periodically strained Co nanomagnet at various bias fields

Figure S2 shows the comparison between the amplitudes of the Kerr oscillation and reflectivity oscillation. The former is two orders of magnitude larger.

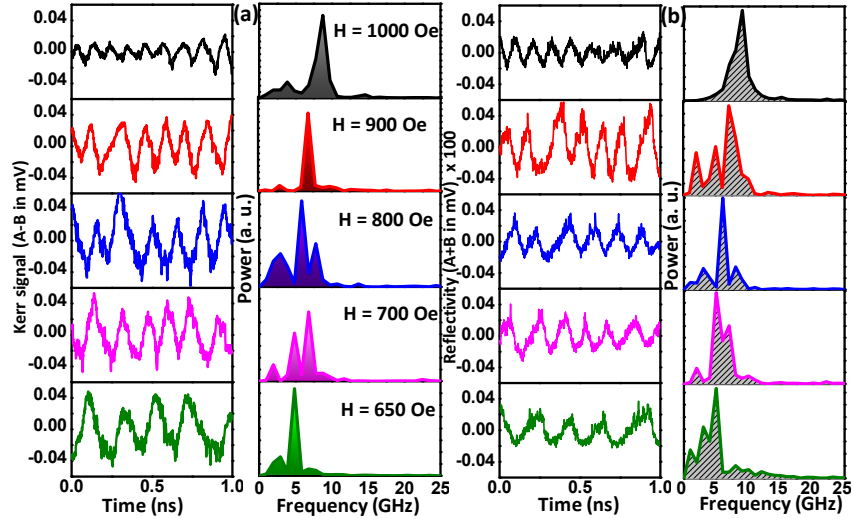

**Figure S2**: Comparison between the amplitudes of the time-resolved (a) Kerr oscillations (A-B) and (b) reflectivity oscillations (A+B) of a periodically strained single Co nanomagnet on a PMN-PT substrate at different bias magnetic fields. The amplitude of the Kerr oscillations is much larger than that of the reflectivity oscillations. The vertical axis of the A+B (reflectivity) signal has been multiplied by a factor of 100 to make the oscillations visible on this scale. The fluences of pump and probe are 15 mJ/cm$^2$ and 2 mJ/cm$^2$, respectively. The comparison between their Fourier spectra is also shown here.

# Time varying SAW-induced strain anisotropy used in the MuMax3 simulation

The SAW-induced periodic strain anisotropy in the nanomagnet is included in the MuMax3 simulation by making the strain anisotropy energy density $K_1(t) = K_0 \left[ \sin(2\pi f_1 t) + \sin(2\pi f_2 t) \right]$, where $f_1$, $f_2$ are the frequencies of the two dominant SAW modes (2 GHz and 8 GHz) observed on the substrate for the pump fluence of 15 mJ/cm$^2$ used in the experiment. Fig. S3 depicts the time variation of the strain anisotropy energy density. The amplitude of the strain oscillation ($K_0$) is a fitting parameter in the MuMax3 simulation. The best match between simulation and experiment is obtained for $K_0 = 22{,}500$ J/m$^3$. This single fitting parameter reproduced our experimental results in the low field regime very well, but not in the high field regime. A single parameter is obviously not adequate to match experiment with simulation over a wide bias field range.

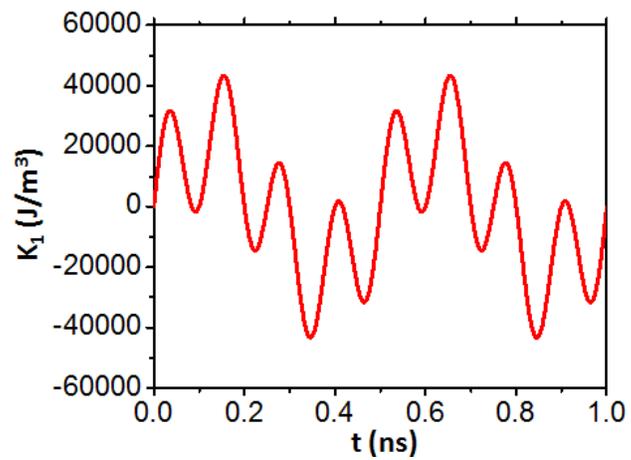

**Figure S3**: Time varying strain anisotropy energy density ($K_1$), which is a combination of two sinusoidal oscillations with frequencies 2 and 8 GHz, is shown here.

# Simulated micromagnetic distribution profile in the periodically strained nanomagnet at various times

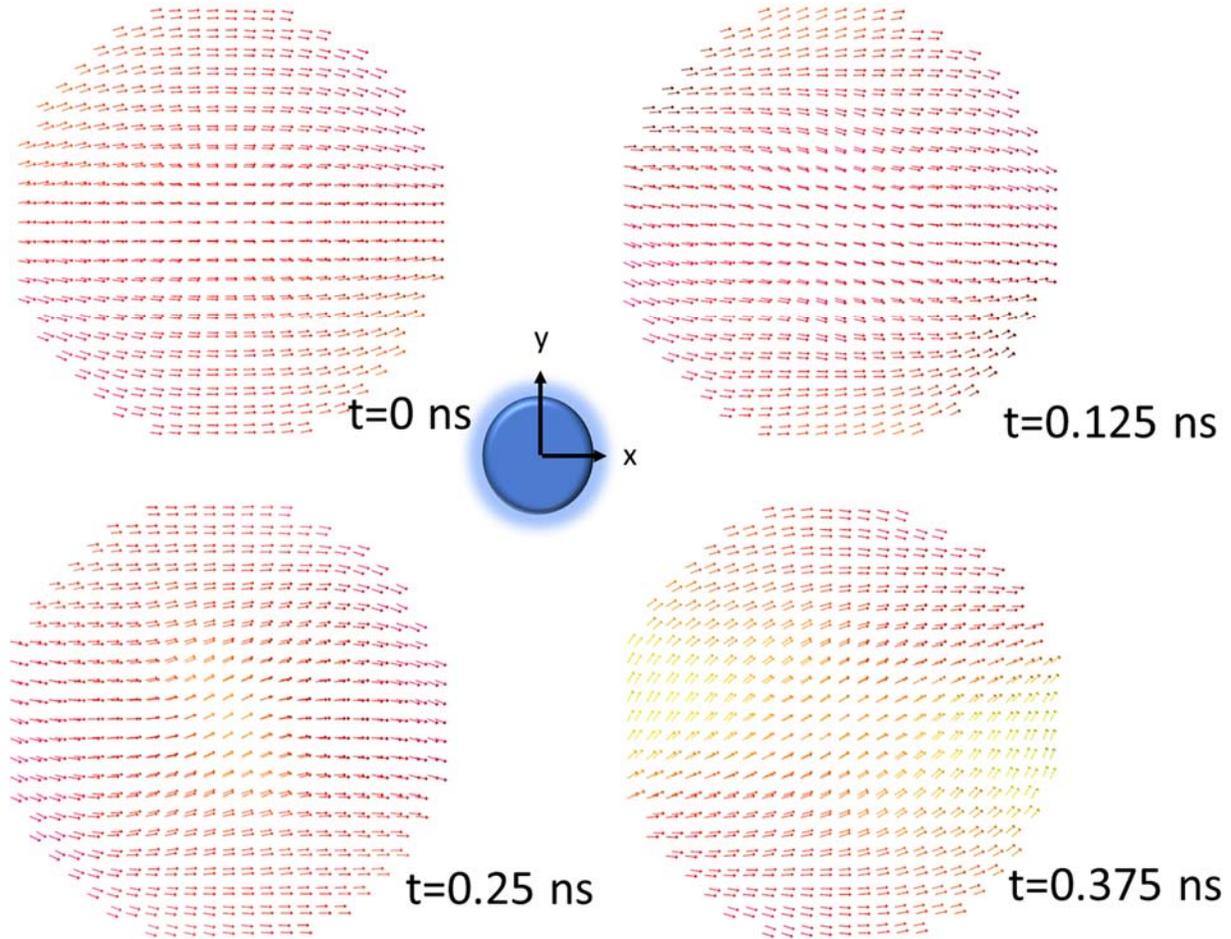

**Figure S4(a)**: Time-lapsed images of the micromagnetic distributions within the Co nanomagnet during the first 0.375 ns obtained from MuMax3 simulations. Time is counted from the instant the time varying strain anisotropy is turned on. The time varying strain anisotropy energy density is again given by $K_1 = 22,500 \left[ \sin(2\pi f_1 t) + \sin(2\pi f_2 t) \right]$; $f_1 = 2$ GHz and $f_2 = 8$ GHz. A bias magnetic field of 760 Oe is directed along the minor axis of the ellipse, pointing to the right. The magnetization is initially assumed to be oriented along the bias field and an out-of-plane square wave tickle field of 30 Oe is turned on at time t = 0 for 100 ps to set the magneto-dynamics in motion.

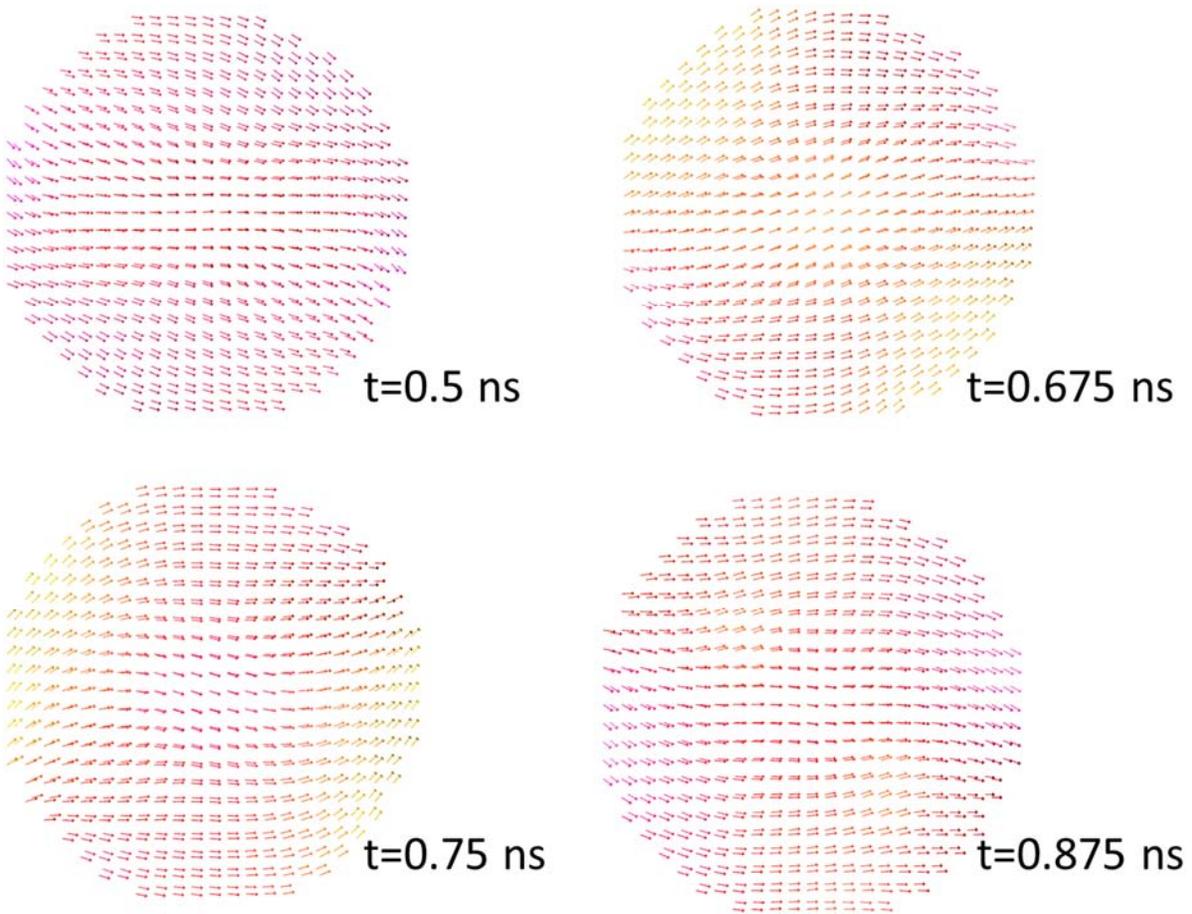

**Figure S4(b):** Time-lapsed images of the micromagnetic distributions within the Co nanomagnet obtained from MuMax3 simulations. The time varying strain anisotropy energy density is again given by $K_1 = 22,500\left[\sin(2\pi f_1 t) + \sin(2\pi f_2 t)\right]$; $f_1 = 2$ GHz and $f_2 = 8$ GHz. A bias magnetic field of 760 Oe is directed along the minor axis of the ellipse, pointing to the right. The magnetization is initially assumed to be oriented along the bias field and an out-of-plane square wave tickle field of 30 Oe is turned on at time t = 0 for 100 ps to set the magneto-dynamics in motion.

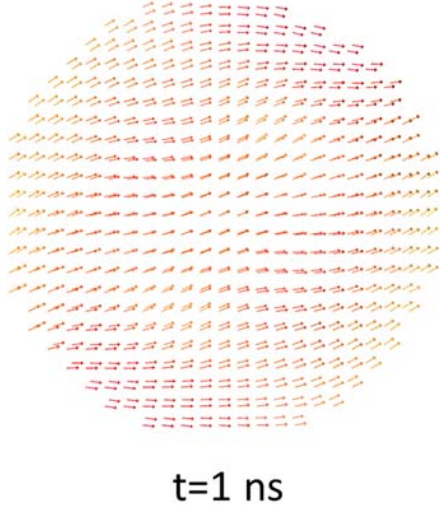

t=1 ns

**Figure S4(c):** Micromagnetic distributions within the Co nanomagnet at 1 ns obtained from MuMax3 simulations. The time varying strain anisotropy energy density is again given by

$K_1 = 22,500\left[\sin(2\pi f_1 t) + \sin(2\pi f_2 t)\right]$; $f_1 = 2$ GHz and $f_2 = 8$ GHz . A bias magnetic field of 760 Oe is directed along the minor axis of the ellipse, pointing to the right. The magnetization is initially assumed to be oriented along the bias field and an out-of-plane tickle field of 30 Oe is turned on at time t = 0 for 100 ps to set the magneto-dynamics in motion.

### Simulated spin wave spectra and corresponding mode profile for the *unstrained* Co nanomagnet

Using MuMax3, we obtained the magnetization dynamics of a single *unstrained* Co nanomagnet (no SAWs) at different bias field values for the purpose of comparison with the periodically strained cases. The same simulation parameters are used as in the periodically strained case, except here $K_0 = 0$. In this case, the magneto-dynamics is governed solely by the laser-induced precession of magnetization around the bias magnetic field.

By fast Fourier transforming the spatially averaged out-of-plane magnetization $\bar{M}_z(t)$, we find that the resulting spectrum has two dominant peaks at any bias magnetic field. These two peaks, for a bias field of 1000 Oe, are shown in Fig. S5(a). The Dotmag software provides the power and phase distributions of the spin waves at these two peak frequencies. They are found to be two standing modes - the center mode (mode 2) and the edge mode (mode 1) for every bias field considered (see Fig. S5(c)). The power distributions of these modes (at the center and vertical edges of the nanomagnet) do not change much with bias field. The bias field dependence of the precessional frequency (Fig. S5(b)) reveals that these two modes have excellent stability. We have fitted the dependence using the Kittel formula,

$$f = \frac{\gamma}{2\pi}\sqrt{(H)(H + 4\pi M_{eff})}$$

where $H$ is the bias magnetic field. The extracted effective magnetization $M_{eff}$ values from the Kittel fit of the field variation of the center- and edge-modes are 1065 emu/cm³ and 325 emu/cm³, respectively. The deviation of $M_{eff}$ values from the intrinsic saturation magnetization of 1100 emu/cm³ (for Co) can be explained by the demagnetizing effect from the unsaturated magnetizations at the nanomagnet edges perpendicular to the bias field direction.

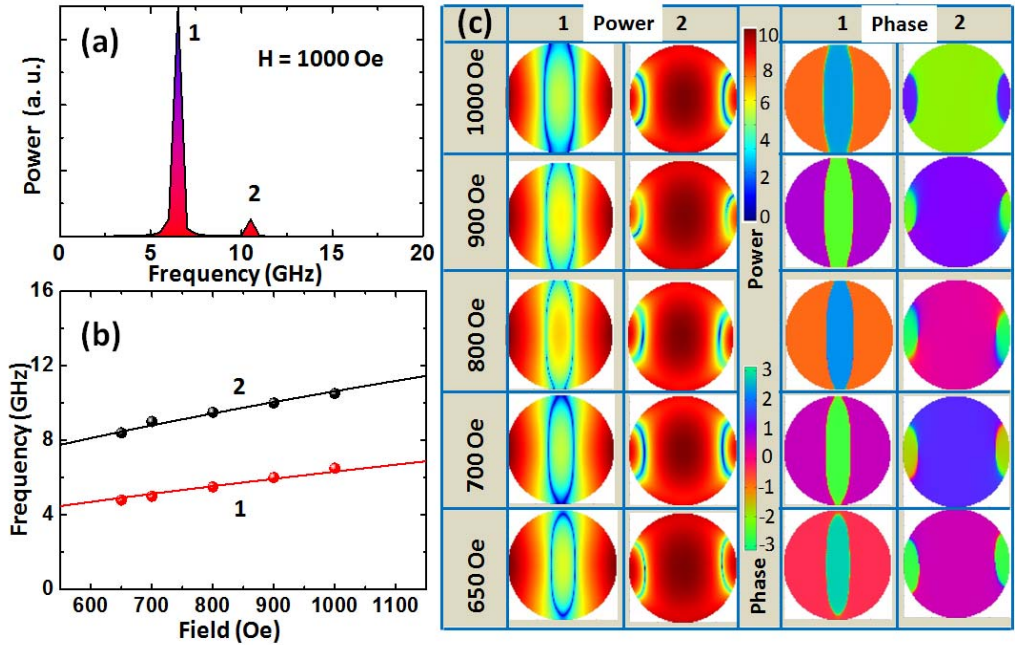

**Figure S5**: (a) Calculated frequency spectrum of the time variation of the spatially-averaged out-of-plane magnetization component $\bar{M}_z(t)$ in an *unstrained* elliptical Co nanomagnet of major axis 190 nm, minor axis 186 nm and thickness 16 nm, in the presence of a bias magnetic field of 1000 Oe directed along the ellipse's minor axis. The quantity $\bar{M}_z(t)$ is calculated from MuMax3. (b) Bias field dependence of the two peak frequencies observed in the calculated frequency spectrum. Solid lines indicate the Kittel fit. (c) Calculated spin-wave mode profile in the nanomagnet for the two peak frequencies in the spectrum at different bias fields. These profiles are calculated with the Dotmag code. Since the power is concentrated at the center in one mode and vertical edges at the other, they are a 'center mode' and an 'edge mode' and this nature is independent of the bias field in this range. The units of power and phase in this plot are dB and radians, respectively.

## A discussion of the two mechanisms that generate SAW in the PMN-PT substrate

There are two mechanisms that generate periodic strain in the PMN-PT substrate and result in a surface acoustic wave – the first is associated with the alternating electric field in the laser generating periodic (compressive and tensile) strain in the PMN-PT substrate from $d_{33}$ and/or $d_{31}$ coupling. The second arises from the (periodic) differential thermal expansions of Co and PMN-PT. Note that the former mechanism requires a piezoelectric substrate, while the latter does not. We will compare the relative strengths of the two.

The nanomagnets are produced by electron beam evaporation of Co through lithographically defined windows on a resist. If, instead, they were a thin film epitaxially grown on the substrate, then we could have assumed that the film is pseudomorphic and in that case, the strain in the nanomagnet due to differential thermal expansion would be approximately

$$\varepsilon = \left(\alpha_{PMN-PT} - \alpha_{Co}\right)\Delta T$$

where $\alpha$-s are the thermal expansion coefficients and $\Delta T$ is the temperature rise. However, the nanomagnets used here are amorphous or polycrystalline and not a pseudomorphic layer. The actual strain is therefore much less in magnitude than what the above equation predicts. Let us assume therefore that

$$\varepsilon_{actual} = \xi\left(\alpha_{PMN-PT} - \alpha_{Co}\right)\Delta T \quad [\xi \ll 1]$$

According to the literature, $\alpha_{Co} = 13 \times 10^{-6}$/K and $\alpha_{PMN-PT} = 9.5 \times 10^{-6}$/K. We estimated in the paper that the strain generated is about 0.18% based on our MuMax3 modeling. If even half of this was due to thermal effects, then the temperature rise needed for that is $257/\xi$ Kelvin (calculated from the last equation). We do not know what $\xi$ is, except that it must be much smaller than unity. Therefore, in order for the thermally generated strain to be a significant fraction (~50%) of the total strain of 0.18%, the temperature rise due to the laser has to be $\gg 257$ K, which is unlikely at the time scale of laser excitation (femtosecond laser) and the applied pump fluence of 15 mJ/cm². Therefore, it stands to reason that the bulk of the strain is generated by the $d_{31}$ and $d_{33}$ coupling of the electric field in the piezoelectric that is modulated by the laser and not by the differential thermal expansion. In other words, the first mechanism is dominant over the second.